\documentclass[amsfonts, amssymb, amsmath, floatfix, superscriptaddress, showpacs, aps, reprint]{revtex4-1}
\usepackage{bm}
\usepackage{graphicx}
\usepackage{epstopdf}

\begin{document}
\title{Physical properties of a candidate quantum spin-ice system \bm{${\rm Pr_2Hf_2O_7}$}}
\author{V.\ K.\ Anand}
\altaffiliation{vivekkranand@gmail.com}
\affiliation{\mbox{Helmholtz-Zentrum Berlin f\"{u}r Materialien und Energie GmbH, Hahn-Meitner Platz 1, D-14109 Berlin, Germany}}
\author{L.\ Opherden}
\affiliation{Dresden High Magnetic Field Laboratory (HLD-EMFL), Helmholtz-Zentrum Dresden-Rossendorf, Bautzner Landstra{\ss}e 400, D-01328 Dresden, Germany}
\author{J.\ Xu}
\affiliation{\mbox{Helmholtz-Zentrum Berlin f\"{u}r Materialien und Energie GmbH, Hahn-Meitner Platz 1, D-14109 Berlin, Germany}}
\affiliation{\mbox{Institut f\"{u}r Festk\"{o}rperphysik, Technische Universit\"{a}t Berlin, Hardenbergstra{\ss}e 36, D-10623 Berlin, Germany}}
\author{D.\ T.\ Adroja}
\affiliation{ISIS Facility, Rutherford Appleton Laboratory, Chilton, Didcot, Oxon, OX11 0QX, United Kingdom}
\affiliation{\mbox{Highly Correlated Matter Research Group, Physics Department, University of Johannesburg, P.O. Box 524,} Auckland Park 2006, South Africa}
\author{\mbox{A.\ T.\ M.\ N.\ Islam}}
\affiliation{\mbox{Helmholtz-Zentrum Berlin f\"{u}r Materialien und Energie GmbH, Hahn-Meitner Platz 1, D-14109 Berlin, Germany}}
\author{T.\ Herrmannsd\"orfer}
\author{J.\ Hornung}
\author{R.\ Sch\"onemann}
\author{M.\ Uhlarz} 
\affiliation{Dresden High Magnetic Field Laboratory (HLD-EMFL), Helmholtz-Zentrum Dresden-Rossendorf, Bautzner Landstra{\ss}e 400, D-01328 Dresden, Germany}
\author{H.\ C.\ Walker}
\affiliation{ISIS Facility, Rutherford Appleton Laboratory, Chilton, Didcot, Oxon, OX11 0QX, United Kingdom} 
\author{N.\ Casati}
\affiliation{\mbox{Paul Scherrer Institute, 5232 Villigen PSI, Switzerland}}
\author{B.\ Lake}
\altaffiliation{bella.lake@helmholtz-berlin.de}
\affiliation{\mbox{Helmholtz-Zentrum Berlin f\"{u}r Materialien und Energie GmbH, Hahn-Meitner Platz 1, D-14109 Berlin, Germany}}
\affiliation{\mbox{Institut f\"{u}r Festk\"{o}rperphysik, Technische Universit\"{a}t Berlin, Hardenbergstra{\ss}e 36, D-10623 Berlin, Germany}}

\date{\today}

\begin{abstract}
Physical properties of a pyrohafnate compound Pr$_2$Hf$_2$O$_7$ have been investigated by ac magnetic susceptibility $\chi_{\rm ac}(T)$, dc magnetic susceptibility  $\chi(T)$, isothermal magnetization $M(H)$ and heat capacity $C_{\rm p}(T)$ measurements on polycrystalline as well as single crystal samples combined with high-resolution synchrotron x-ray diffraction (XRD) for structural characterization and inelastic neutron scattering (INS) to determine the crystal field energy level scheme and wave functions. Synchrotron XRD data confirm the ordered cubic pyrochlore ($Fd\bar3m$) structure without any noticeable site mixing or oxygen deficiency. No clear evidence of long range magnetic ordering is observed down to 90~mK, however the $\chi_{\rm ac}(T)$ evinces slow spin dynamics revealed by a frequency dependent broad peak associated with spin freezing. The INS data reveal the expected five well defined magnetic excitations due to crystal field splitting of the $J=4$ ground state multiplet of the Pr$^{3+}$. The crystal field parameters and ground state wavefunction of Pr$^{3+}$ have been determined. The Ising anisotropic nature of magnetic ground state is inferred from the INS as well as $\chi(T)$ and $M(H)$ data. Together these properties make Pr$_2$Hf$_2$O$_7$ a candidate compound for quantum spin-ice behavior. 
\end{abstract}


\maketitle

\section{\label{Intro} INTRODUCTION}

The rare earth pyrochlore oxides $R_2B_2$O$_7$ ($R$ is a trivalent rare earth and $B$ a tetravalent transition metal or Ge, Sn, Pb) containing magnetic networks of corner-sharing tetrahedra are well known for their exotic ground state (arising from the combined effect of crystal field anisotropy and dipolar and exchange interactions) and have been an interesting topic of research for the past few decades \cite{Gardner2010, Castelnovo2012, Gingras2014, Ramirez1999, Siddharthan1999,Hertog2000,Bramwell2001}. The research interests in these systems were invigorated after the discovery of spin-ice behavior in Dy$_2$Ti$_2$O$_7$ and Ho$_2$Ti$_2$O$_7$ \cite{Ramirez1999, Siddharthan1999, Hertog2000, Bramwell2001}. The crystal field induced Ising anisotropic ground state (where the spins are forced to point along the local $\langle 111 \rangle $ direction) of these classical spin-ice materials along with ferromagnetic dipolar interactions give rise to a `two-in/two-out' configuration (the so called `ice rule') of rare earth moments \cite{Harris1997}. The two-in/two-out spin configuration of these classical spin-ices is equivalent to the arrangement of protons in water ice, thereby they also possess low-temperature Pauling residual entropy $(1/2) R \ln(3/2)$ \cite{Ramirez1999}. The fundamental excitations at the heart of spin-ice physics are the emergent magnetic monopoles arising from the creation and anihilation of `three-in/one-out' or `one-in/three-out' spin configurations \cite{Castelnovo2008}. The experimental observation of Dirac strings via neutron scattering have provided evidence of these magnetic monopoles \cite{Castelnovo2008, Morris2009, Fennell2009}. The monopole dynamics provides further insights to the emergent magnetic phases in spin-ice materials \cite{Castelnovo2008, Fennell2009, Morris2009, Jaubert2009, Bramwell2009}. 

The titanate pyrochlores Tb$_2$Ti$_2$O$_7$ and Yb$_2$Ti$_2$O$_7$ also present interesting magnetic ground states. Tb$_2$Ti$_2$O$_7$ is a spin-liquid system that shows a cooperative critical paramagnetic ground state \cite{Gardner1999, Gardner2001, Molavian2007}. Yb$_2$Ti$_2$O$_7$ exhibits the characteristic features of quantum spin-ice (QSI) behavior \cite{Ross2011, Pan2016}. The ground state of Yb$_2$Ti$_2$O$_7$ is however very much sample dependent ranging from ferromagnetic order to disorder. A Higgs transition from a magnetic Coulomb liquid state to a ferromagnetic state (the Higgs phase of magnetic monopoles) has been suggested for ferromagnetically ordered Yb$_2$Ti$_2$O$_7$ \cite{Chang2012}. The quantum fluctuations play an important role in quantum spin-ice materials. While in a classical spin-ice system, because of strong magnetic anisotropy, only Ising interactions exist, in a quantum spin-ice system the magnetic anisotropy is not so strong and significant non-Ising terms exist in the Hamiltonian, therefore quantum fluctuations strongly dominate the spin dynamics in QSI systems. In order to have strong quantum fluctuations the dipolar interaction has to be small which one would naively expect to be the case with the materials having rare earth ions with smaller moments. This condition is fulfilled for Yb$^{3+}$, and because of the smaller dipolar interaction the ground state of Yb$_2$Ti$_2$O$_7$ is mainly dominated by exchange interactions \cite{Ross2011}. 

Pr$^{3+}$ and Ce$^{3+}$ have much smaller magnetic moments compared to Dy$^{3+}$ and Ho$^{3+}$, therefore the pyrochlore systems containing Pr and Ce is expected to fullfill the condition of smaller dipolar interaction. As the dipolar interaction is proportional to the square of magnetic moment, the dipolar interaction in Pr-pyrochlores (say Pr$_2$Sn$_2$O$_7$)  will be only  $\sim 9\%$ of that in Dy$_2$Ti$_2$O$_7$. Indeed the Pr-pyrochlores Pr$_2$Sn$_2$O$_7$ \cite{Zhou2008} and Pr$_2$Zr$_2$O$_7$ \cite{Kimura2013}, and Ce-pyrochlore Ce$_2$Sn$_2$O$_7$ \cite{Sibille2015} have been found to show the characteristic features of a quantum spin-ice/spin-liquid system. Pr$_2$Ir$_2$O$_7$ on the other hand shows a metallic spin-liquid ground state \cite{Nakatsuji2006} that is situated close to a quantum critical point \cite{Tokiwa2014}. Of particular interest are the Pr-based pyrochlores which have a non-Kramers doublet as ground state. Lee {\it et al}.\ \cite{Lee2012} used a gauge mean-field theoretical approach to suggest that the QSI phase in a system with a non-Kramers doublet ground state [such as in Pr$_2B_2$O$_7$  ($B$ = Zr, Sn, Ir)]  is more stable than in a system with a Kramers doublet ground state. 

The hafnate pyrochlore Pr$_2$Hf$_2$O$_7$ is another potential Pr-based system for the study of possible quantum spin-ice behavior. Pr$_2$Hf$_2$O$_7$ is reported to form in a cubic pyrochlore ($Fd\bar3m$) structure \cite{Karthik2012}. A recent muon spin relaxation study on Pr$_2$Hf$_2$O$_7$ by Foronda {\it et al}.\ \cite{Foronda2015} found muon-induced anisotropic local distortion for which they proposed a splitting of the non-Kramers doublet of Pr$^{3+}$. Extending our work on pyrohafnates \cite{Anand2015} here we report the physical properties of Pr$_2$Hf$_2$O$_7$ which indeed seems to be a potential candidate for quantum spin-ice behavior. Our investigations using various complementary techniques exclude long-range magnetic ordering down to 90~mK\@.  The ac magnetic susceptibility $\chi_{\rm ac}(T)$ shows a rather broad frequency dependent anomaly near 0.2~K at 160~Hz. The frequency dependent behavior of this anomaly at such a low temperature is a reminiscence of spin-ice dynamics. With a $J$ value of 4, Pr$^{3+}$ has a $(2J+1 = 9)$-fold degenerate ground state which under the action of crystal electric field (CEF) can split into a combination of singlets and doublets. In order to determine whether the CEF-split ground state of Pr$^{3+}$ is consistent with spin-ice physics we performed an inelastic neutron scattering measurement and found that Pr$_2$Hf$_2$O$_7$ indeed has a well isolated Ising anisotropic doublet ground state. The CEF ground state of Pr$_2$Hf$_2$O$_7$ thus meets the requirements for quantum spin-ice behavior. A recent work by Sibille {\it et al}. \cite{Sibille2016} also reports a quantum spin-ice behavior in Pr$_2$Hf$_2$O$_7$. 

\section{\label{ExpDetails} EXPERIMENTAL DETAILS}

The polycrystalline Pr$_2$Hf$_2$O$_7$ sample was prepared at the Crystal Lab, Helmholtz-Zentrum Berlin (HZB) by solid-state reaction method using the high purity materials: Pr$_6$O$_{11}$ (99.99\%, Alfa Aesar)  and HfO$_2$ (99.95\%, Alfa Aesar). A single phase sample was obtained by firing the stoichiometric mixture of Pr$_6$O$_{11}$ and HfO$_2$ at 1300~$^\circ$C for 50~h followed by three more successive grindings, pelletizings and firings at 1500~$^\circ$C for 50~h\@. The polycrystalline La$_2$Hf$_2$O$_7$ sample was prepared using La$_2$O$_3$ (99.999\%, Alfa Aesar) and HfO$_2$  as detailed in Ref.~\cite{Anand2015}. The phase purity and quality of the samples were checked by room temperature powder x-ray diffraction using the Bruker-D8 laboratory-based x-ray diffractometer. 

A large single crystal of Pr$_2$Hf$_2$O$_7$ was successfully grown by the Floating-zone technique in a four mirror type high-temperature optical floating zone furnace (Crystal Systems corp.\ FZ-T-12000-X-VII-VPO-PC) equipped with Xenon arc lamps, in flowing argon atmosphere. The quality of the single crystal was checked by Laue diffraction. The single crystal was also oriented by the Laue method and cut into small pieces for magnetic measurements.

The magnetic measurements down to 2~K and up to 14~T were performed using a Quantum Design superconducting quantum interference device vibrating sample magnetometer (SQUID-VSM) and VSM option of a Quantum Design physical properties measurement system (PPMS) at Mag Lab, HZB. The heat capacity measurements down to 1.8~K were performed by the adiabatic-relaxation technique using the PPMS at Mag Lab, HZB.  The ac susceptibility measurements down to 90~mK were performed  at Helmholtz-Zentrum Dresden-Rossendorf using compensated coil-pair susceptometer built in a He-3/4 dilution refrigerator using a SR830 lock-in amplifier. We also made an attempt to measure the resistivity but the resistance was found to be beyond the measurement range of PPMS indicating an insulating behavior. The resistance measurement using a multimeter further confirmed the insulating ground state of Pr$_2$Hf$_2$O$_7$.

The high-resolution synchrotron x-ray diffraction measurements were performed on the MS-beamline \cite{Willmott2013} at Paul Scherrer Institute (PSI), Switzerland. The sample was very finely ground and filled into a thin capillary (0.3~mm in diameter).  In order to minimize the profile shape dependence on capillary diameter and reduce the effect of preferred orientation the sample was rotated continuously. X-rays of energy 25~keV were used for the measurement. The exact wavelength of the x-rays (0.4957 {\AA}) and the instrument profile parameters were determined by measurements on the standard LaB$_6$ powder (NIST) under the identical experimental condition. 

The inelastic neutron scattering measurements were performed on the high neutron flux beamline MERLIN at the ISIS facility of the Rutherford Appleton Laboratory, Didcot, U.K. \cite{Bewley2006, Bewley2009}. About 20~g powder samples each of Pr$_2$Hf$_2$O$_7$ and La$_2$Hf$_2$O$_7$ were used for the INS measurements. The powder samples were mounted inside a thin-walled aluminum can using a thin wrapper of aluminum foil in a cylindrical geometry with diameter approximately 40~mm and height 40~mm, which was then mounted into a closed-cycle refrigerator using \mbox{He-4} exchange gas. The measurements were carried out at 10~K, 100~K and 300~K using neutrons of incident energies $E_i = 18.9$~meV, 40.1~meV, 86.3~meV, 136.0~meV and 195.0~meV.

\section{\label{Structure} Crystallography}

\begin{figure}
\includegraphics[width=\columnwidth, keepaspectratio]{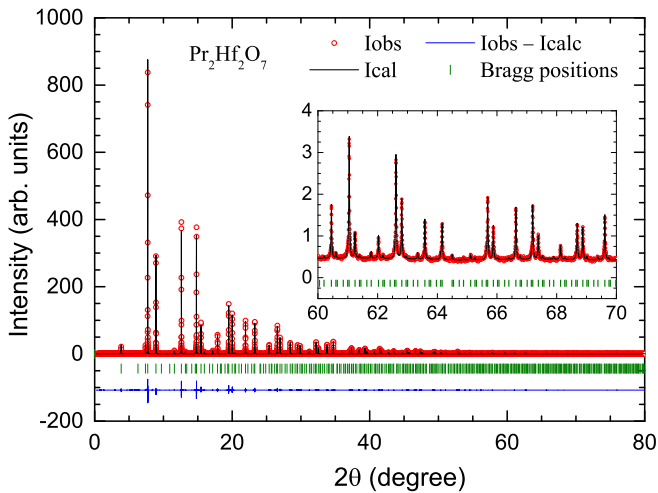}
\caption {(Color online) Synchrotron powder x-ray diffraction pattern of polycrystalline Pr$_2$Hf$_2$O$_7$ recorded at room temperature. The solid line through the experimental points is the Rietveld refinement profile calculated for the ${\rm Eu_2Zr_2O_7}$-type face-centered cubic (space group $Fd\bar{3}m$) pyrochlore structure. The short vertical bars mark the Bragg peaks positions and the lowermost curve represents the difference between the experimental and calculated intensities. Inset: An expanded scale view showing the details of the refinement in a small angle range at high $2\theta$.}
\label{fig:XRD}
\end{figure}

The room temperature high-resolution synchrotron powder XRD pattern of polycrystalline Pr$_2$Hf$_2$O$_7$ is shown in Fig.~\ref{fig:XRD} along with the structural Rietveld refinement profile. For Rietveld refinement we used the software FullProf \cite{Rodriguez1993}. The refinement revealed no impurity phase and confirmed the reported ${\rm Eu_2Zr_2O_7}$-type face-centered cubic  (space group $Fd\bar{3}m$) pyrochlore structure. In this structure both Pr and Hf form individual three-dimensional networks of corner-shared tetrahedra. The Wyckoff positions (atomic coordinates) of Pr, Hf, O1 and O2 are 16d (1/2,1/2,1/2), 16c (0,0,0), 48f ($x_{\rm O1}$,1/8,1/8) and 8b (3/8,3/8,3/8), respectively. The lattice parameter $a = 10.6728(1)$~\AA\ and the $x$-coordinate of O1 $x_{\rm O1} = 0.3351(4)$ obtained from the refinement agree very well with the reported values \cite{Karthik2012}. The agreement factors for the refinement are $\chi^2 = 10.1$, $R_{\rm p} = 2.91\%$ and $R_{\rm wp} = 3.66\%$. We also checked for possible Pr/Hf site mixing and refined the occupancies of atoms which did not make any noticeable improvement in the fit or agreement factors. From our synchrotron data we anticipate that Pr/Hf site mixing and/or oxygen deficiency, if present, are less than 0.5\%. This suggests a well-ordered pyrochlore structure for Pr$_2$Hf$_2$O$_7$ consistent with the previous report  \cite{Karthik2012} and with the prediction based on the ratio of the cation radii $r_{\rm Pr}/r_{\rm Hf} \approx 1.59 $ which is well within the range of 1.36--1.71 required for a stable ordered pyrochlore phase  \cite{Gardner2010}.

\section{\label{Sec:ChiMH} DC Magnetic Susceptibility and Magnetization}

\begin{figure}
\includegraphics[width=\columnwidth, keepaspectratio]{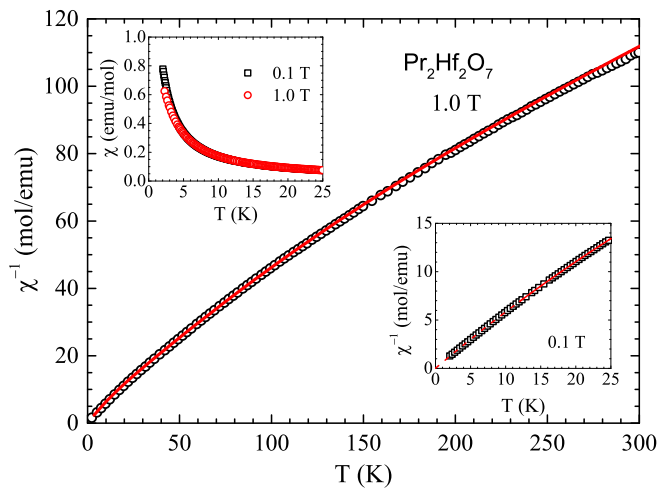}
\caption{(Color online) Zero-field-cooled magnetic susceptibility $\chi$ of polycrystalline Pr$_2$Hf$_2$O$_7$ plotted as inverse magnetic susceptibility $\chi^{-1}$ as a function of temperature $T$ for $2~{\rm K} \leq T \leq 300$~K measured in magnetic field $H= 1.0$~T\@. The solid curve is the crystal field susceptibility corresponding to the crystal field parameters obtained from the analysis of inelastic neutron scattering data. Upper inset: Low-$T$ $\chi(T)$ for $H= 0.1$~T and 1.0~T and $2~{\rm K} \leq T \leq 25$~K\@. Lower inset: $\chi^{-1}(T)$ for $H=0.1$~T and $2~{\rm K} \le T \leq 25$~K\@. The solid red line is the fit of the $\chi^{-1}(T)$ data by the modified Curie-Weiss law fitted over $12~{\rm K} \leq T \leq 25$~K and the dashed line is an extrapolation.}
\label{fig:MT}
\end{figure}

The zero-field-cooled (ZFC) dc magnetic susceptibility $\chi(T)$ data of polycrystalline Pr$_2$Hf$_2$O$_7$ are shown in Fig.~\ref{fig:MT}. It is clear from Fig.~\ref{fig:MT} that at $T \geq 2$~K the $\chi(T)$ data show no anomaly related to magnetic phase transition. As can be seen from the upper inset of Fig.~\ref{fig:MT}, the magnitude of $\chi$ at low-$T$ is rather large. The ZFC and field-cooled (FC) $\chi(T)$ data (not shown) do not show any thermal hysteresis above 2~K\@. The $\chi(T)$ data follow modified Curie-Weiss behavior, $\chi(T) =  \chi_0 + C/(T-\theta_{\rm p})$ (lower inset of Fig.~\ref{fig:MT}). A fit of ZFC $\chi^{-1}(T)$ data (measured in magnetic field $H=0.1$~T) in $12~{\rm K} \leq T \leq 25$~K gives $\chi_0 = 5.02(6) \times 10^{-3}$ emu/mol\,Pr, $C = 0.81(1)$~emu\,K/mol\,Pr and $\theta_{\rm p} = -0.01(2)$~K\@. Upon correcting the $\chi(T)$ data for demagnetization effects (approximating the sample by a sphere) we obtain $\chi_0 = 5.02(6) \times 10^{-3}$ emu/mol\,Pr, $C = 0.81(1)$~emu\,K/mol\,Pr and $\theta_{\rm p} = +0.06(2)$~K\@. The positive $\theta_{\rm p}$ suggests the effective interaction in Pr$_2$Hf$_2$O$_7$ to be ferromagnetic in nature. The $C$ value yields an effective moment of $\mu_{\rm eff} \approx 2.54\, \mu_{\rm B}$/Pr which is significantly lower than the expected paramagnetic state value of $\mu_{\rm eff} = g_J\sqrt{J(J+1)} = 3.58 \, \mu_{\rm B}$/Pr for free Pr$^{3+}$ ions ($g_J = 4/5$ and $J = 4$). The reduced value of $\mu_{\rm eff}$ reflects the Ising anisotropic nature of magnetic ground state. A fit of the $H=1.0$~T $\chi^{-1}(T)$ data in $50~{\rm K} \leq T \leq 300$~K yields $\mu_{\rm eff}  \approx 2.91 \, \mu_{\rm B}$/Pr which is also smaller than the free ion value and can be attributed to crystal field effect.

\begin{figure}
\includegraphics[width=\columnwidth, keepaspectratio]{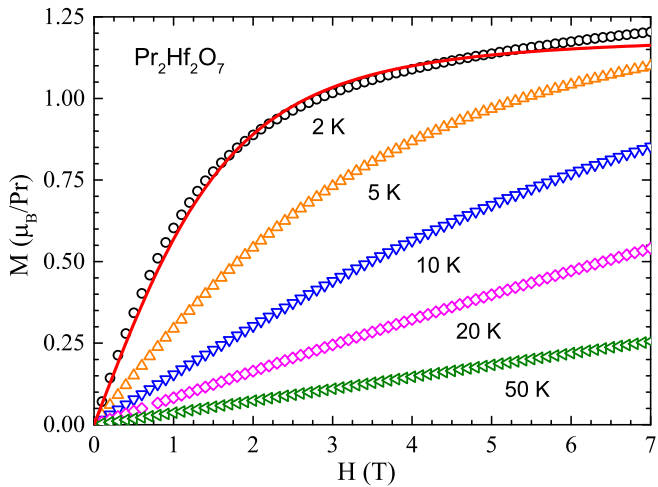}
\caption{(Color online) Isothermal magnetization $M$ of polycrystalline Pr$_2$Hf$_2$O$_7$ as a function of applied magnetic field $H$ for $0 \leq H \leq 7$~T measured at the indicated temperatures. The solid curve is the fit of 2~K $M(H)$ data by Eq.~(\ref{MH-Ising}) with an effective longitudinal $g$-factor $g_{\parallel}= 4.78(1)$.}
\label{fig:MH}
\end{figure}

The isothermal magnetization $M(H)$ data of polycrystalline Pr$_2$Hf$_2$O$_7$ at five different temperatures between 2~K and 50~K are shown in Fig.~\ref{fig:MH}. For $T\geq 20$~K the $M(H)$ isotherms are almost linear in $H$ and become nonlinear at $T\leq 10$~K\@. The nonlinearity can be attributed to the saturation behavior of magnetization which is best reflected in $M(H)$ curve at 2~K\@. At 2~K, the near saturation value of magnetization is $M \approx 1.15\, \mu_{\rm B}$/Pr at 5~T which is  only about 36\% of the theoretical saturation magnetization $M_{\rm sat} = g_J J\,\mu_{\rm B} = 3.2\,\mu_{\rm B}$/Pr for free Pr$^{3+}$ ions. However, the observed $M$ is consistent with the local $\langle 111 \rangle$ Ising anisotropic behavior. We fitted the $M(H)$ data by \cite{Bramwell2000, Xu2015}
\begin{equation}
\langle M \rangle = \frac{(k_{\rm B} T)^2}{g_{\parallel}\mu_{\rm B} H^2 S} \int_0^{g_{\parallel}\mu_{\rm B} H S/k_{\rm B} T} x \tanh(x) \,dx
\label{MH-Ising}
\end{equation}
that describes the $H$ dependence of powder-averaged $M$ in the paramagnetic state of an Ising pyrochlore within effective spin $S = 1/2$ model. Here $g_{\parallel}$ is the longitudinal $g$-factor and the transverse $g$-factor $g_\perp$ is assumed to be zero. The fit of $M(H)$ data at 2~K (shown by solid red line in Fig.~\ref{fig:MH}) gives $g_{\parallel} = 4.78(1)$, accordingly the paramagnetic state moment is expected to be $m = g_{\parallel} S\,\mu_{\rm B} \approx 2.4\,\mu_{\rm B}$/Pr which is very close to the $\mu_{\rm eff} \approx 2.54\, \mu_{\rm B}$/Pr obtained above from $\chi(T)$.  Further it is seen that while Eq.~(\ref{MH-Ising}) predicts a saturation tendency at high $H$ (see the fit in Fig.~\ref{fig:MH}), the $M(H)$ data show a weak increase, this departure is clearly noticed at $H> 6$~T\@. Such a behavior of $M(H)$ hints for the presence of spin fluctuations/non-Ising contribution.

The nearest neighbor dipole-dipole interaction $D_{\rm nn}$ corresponding to our Ising ground state moment $\mu_{\rm eff} = 2.54\, \mu_{\rm B}$/Pr can be estimated using the relation $D_{\rm nn} = ({5}/{3})({\mu_0}/{4\pi}) {\mu_{\rm eff}^2}/{r_{\rm nn}^3}$ \cite{Bramwell2001,Hertog2000}, $\mu_0$ being the magnetic permeability of free space and $r_{\rm nn }= (a/4)\surd 2$ the nearest neighbor distance. For Pr$_2$Hf$_2$O$_7$ the lattice parameter $a = 10.6389(1)$~\AA, therefore $D_{\rm nn} \approx 0.12$~K\@. An estimate of effective magnetic interaction $J_{\rm eff}$ can be obtained from the value of Weiss temperature, $\theta_{\rm p} = z J_{\rm eff} S(S+1)/3k_{B}$ ($z=6$ is the number of the nearest neighbor) which for $\theta_{\rm p} = +0.06$~K and $S=1/2$ (effective spin) yields $J_{\rm eff} = +0.04$~K (in our notation a positive sign indicates ferromagnetic interaction and a negative sign the antiferromagnetic interaction). The effective interaction $J_{\rm eff} = J_{\rm nn} + D_{\rm nn}$ where $J_{\rm nn}$ is the nearest neighbor exchange interaction which for $J_{\rm eff} = +0.04$~K and $D_{\rm nn} \approx +0.12$~K would imply $J_{\rm nn} \approx -0.08$~K\@. As expected, the exchange interaction is thus antiferromagnetic in nature. A more realistic estimate of $J_{\rm nn}$ can be obtained from heat capacity $C_{\rm p}(T)$. The height or the temperature of the peak of $C_{\rm p}(T)$ has been found to give very reasonable estimate of $J_{\rm nn}$ \cite{Gardner2010,Hertog2000,Melko2004} which does not suffer from the issue of choice of fitting range of temperature that causes a little variation in the value of $\theta_{\rm p}$. Sibille {\it et al}. \cite{Sibille2016} reported a value of $C_{\rm p} \approx 3.2$~J/mol\,K near the peak around 1.8~K\@.  The $C_{\rm peak} \approx 3.2$~J/mol\,K corresponds to $J_{\rm nn}/D_{\rm nn} \approx 0.7$ in the phase diagram of dipolar spin ice  \cite{Hertog2000} which for $D_{\rm nn} \approx +0.12$~K yields $J_{\rm nn} \approx -0.08$~K\@. We thus see that both $\theta_{\rm p} $ and $C_{\rm peak}$ give a very consistent estimate of the antiferromagnetic exchange interaction in Pr$_2$Hf$_2$O$_7$. 

\section{\label{Sec:ACsus} AC Magnetic Susceptibility}

\begin{figure*}
\includegraphics[width=6.5in, keepaspectratio]{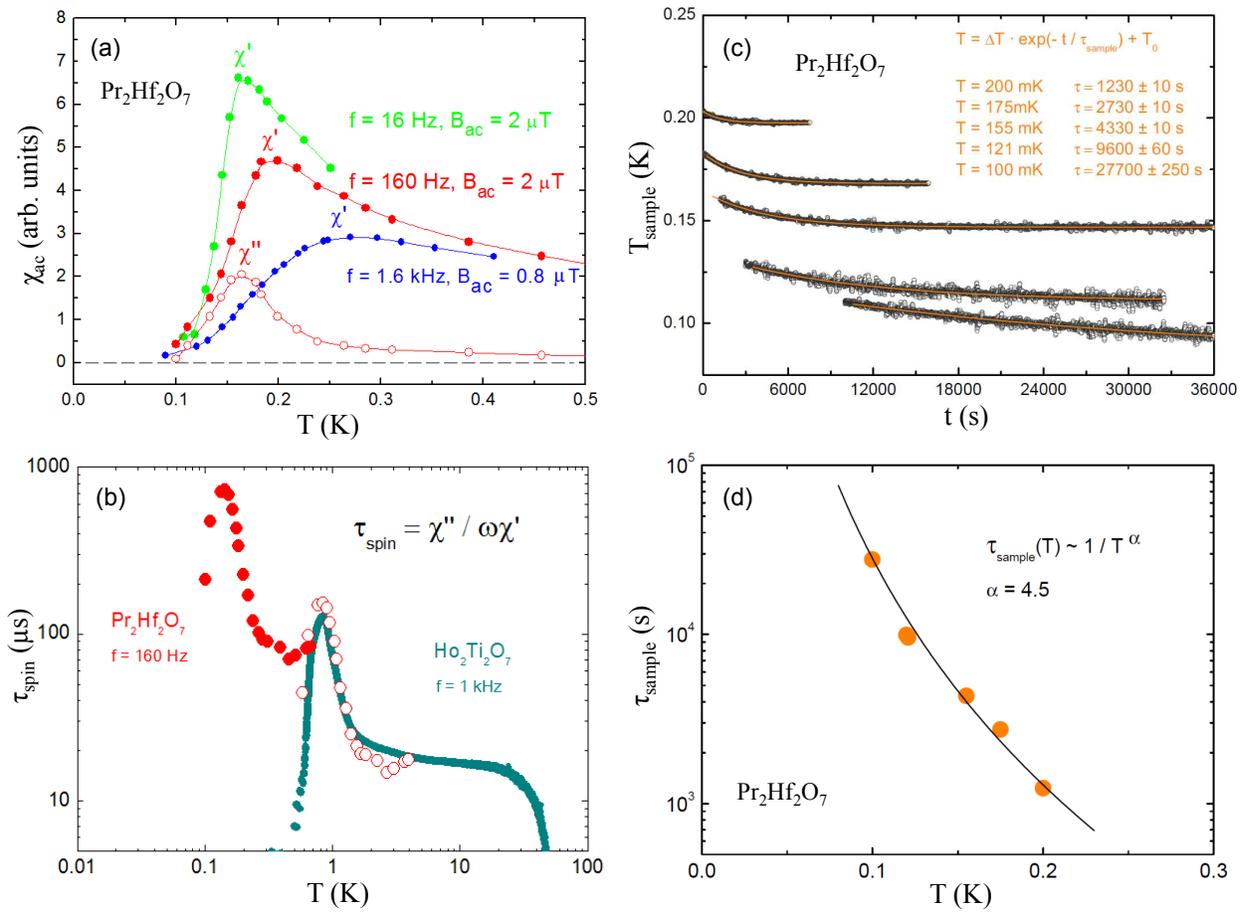}
\caption {(Color online) (a) Real  $\chi'$ and imaginary $\chi''$  parts of zero-field-cooled ac magnetic susceptibility $\chi_{\rm ac}$ of polycrystalline Pr$_2$Hf$_2$O$_7$ as a function of temperature $T$ for $0.09~{\rm K} \leq T \leq 0.5$~K measured at indicated frequencies and ac magnetic fields (in zero dc field). (b) Single-spin relaxation time $\tau_{\rm spin}(T) = \chi''(T)/\omega  \chi'(T)$ of Pr$_2$Hf$_2$O$_7$. Also shown is the $\tau_{\rm spin} (T)$ for Ho$_2$Ti$_2$O$_7$ \cite{Thomas}. The open circles are the data points of Pr$_2$Hf$_2$O$_7$ normalized and scaled to the spin-freezing temperature of Ho$_2$Ti$_2$O$_7$. (c) Sample temperature $T_{\rm sample}$ as a function of time $t$ showing the thermal relaxation in polycrystalline Pr$_2$Hf$_2$O$_7$. The orange solid  lines are the fits of the $T_{\rm sample}(t)$ data. (d) Thermal relaxation time $\tau_{\rm sample}(T)$ obtained from the fitting of $T_{\rm sample}(t)$ in (c).  } 
\label{fig:Chiac}
\end{figure*}

The ac magnetic susceptibility $\chi_{\rm ac}$ of polycrystalline Pr$_2$Hf$_2$O$_7$ measured at frequencies 16~Hz, 160~Hz and 1.6~kHz is shown in Fig.~\ref{fig:Chiac}(a). Both real $\chi'$ and imaginary $\chi''$ parts of $\chi_{\rm ac}(T)$ show broad peaks. The peak position depends on the frequency, with increasing frequency the peak position shifts to higher temperatures. At 16~Hz $\chi'(T)$ shows a peak near 0.16~K which shifts to 0.18 K at 160 Hz and 0.28~K at 1.6~kHz. The frequency dependent ac susceptibility clearly indicates very slow spin dynamics and is reminiscent of spin-ice behavior. This kind of frequency dependent $\chi_{\rm ac}(T)$ associated with spin-freezing has also been observed in the classical spin-ice systems Dy$_2$Ti$_2$O$_7$  \cite{Matsuhira2001} and Ho$_2$Ti$_2$O$_7$ \cite{Matsuhira2000} and dynamic spin-ice systems Pr$_2$Sn$_2$O$_7$ \cite{Matsuhira2004} and Pr$_2$Zr$_2$O$_7$ \cite{Kimura2013}. Sibille {\it et al}.\ [29]  also found a similar frequency dependent peak in ac susceptibility of Pr$_2$Hf$_2$O$_7$. Through the $\chi_{\rm ac}$ measurements at several frequencies they suggested cooperative spin-dynamics in this compound.

We have determined the single-spin relaxation time $\tau_{\rm spin}$ according to the Casimir-du Pr\'e relation in the paramagnetic limit \cite{Casimir1938}, $\tau_{\rm spin} = \chi''/\omega  \chi'$, the $T$ dependence of which is shown in Fig.~\ref{fig:Chiac}(b). For comparison, $\tau_{\rm spin} (T)$ of Ho$_2$Ti$_2$O$_7$ \cite{Thomas} is also shown in Fig.~\ref{fig:Chiac}(b). The single-spin relaxation time is a sensitive and direct measure of the spin dynamics of magnetic materials, in particular in the vicinity of the spin-freezing temperature. The $\tau_{\rm spin} (T)$ of Pr$^{3+}$ magnetic moments of Pr$_2$Hf$_2$O$_7$ scales pretty well with that of Ho$^{3+}$ in Ho$_2$Ti$_2$O$_7$ by multiplying with the ratio of their spin-freezing temperatures. We thus see that $\tau_{\rm spin} (T)$ of Pr$_2$Hf$_2$O$_7$ follows qualitatively and quantitatively the temperature dependent single-spin relaxation time observed for the spin-ice compound Ho$_2$Ti$_2$O$_7$.

We also see a distinct thermal relaxation behavior in Pr$_2$Hf$_2$O$_7$. However, the ac susceptibility of the sample has a reproducible temperature dependence and serves as an internal sample thermometer. Susceptibility data have been taken during stepwise heating up and cooling of the mixing chamber of the used He-3/4 fridge over long periods. At very low temperatures, the temperature has been kept constant for days [relaxed susceptibility-temperature data points are shown in Fig.~\ref{fig:Chiac}(a)]. 

The sample temperature $T_{\rm sample}$ deduced from the $\chi_{\rm ac}$ data is shown in Fig.~\ref{fig:Chiac}(c) as a function of time $t$. The $T_{\rm sample}(t)$ represents the thermal relaxation in polycrystalline Pr$_2$Hf$_2$O$_7$. The $T_{\rm sample}(t)$ data in Fig.~\ref{fig:Chiac}(c) show a very slow thermal relaxation in Pr$_2$Hf$_2$O$_7$. We fit the $T_{\rm sample}(t)$ data by $T_{\rm sample}(t) = T_0 + \Delta T \exp(-t/\tau_{\rm sample})$, the fits are shown by orange solid  lines in Fig.~\ref{fig:Chiac}(c). The mean thermal relaxation time $\tau_{\rm sample}(T)$ obtained from the fit is shown in Fig.~\ref{fig:Chiac}(d). The $\tau_{\rm sample}$ is very large and increases very rapidly as $T$ decreases, following a power law behavior, $\tau_{\rm sample}(T) = R_{\rm thermal}(T) C_{\rm sample}(T) \sim 1/T^\alpha$ with $\alpha=4.5$. The observed thermal relaxation time of the sample exceeds the typical time windows of standard heat-capacity measurements. As the thermal coupling resistance $R_{\rm thermal}(T) $ is expected to follow a $1/T^3$ power law due to phononic transport in the low-temperature limit, the sample relaxation time of Pr$_2$Hf$_2$O$_7$ allows for an estimate of the temperature dependence of the heat capacity, $C_{\rm sample}(T) \sim 1/ T^{1.5}$ in the investigated temperature interval (0.1--0.25~K). This increase of heat capacity with decreasing temperature at low-$T$ indicates that, in contrast to observations in classical spin-ice materials, the spin entropy drops in the temperature range of spin freezing. In consequence, the ice entropy might not be conserved in Pr$_2$Hf$_2$O$_7$. We interpret this observation as a hint for the occurence of quantum fluctuations in the low-temperature limit.

\section{\label{Sec:HC}Heat Capacity}

\begin{figure}
\includegraphics[width=\columnwidth, keepaspectratio]{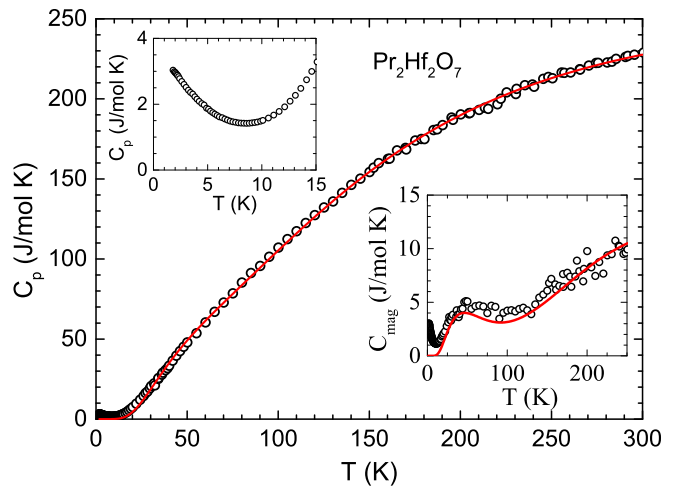}
\caption{\label{fig:HC} (Color online) (a) Heat capacity $C_{\rm p}$ of polycrystalline Pr$_2$Hf$_2$O$_7$ as a function of temperature $T$ for 1.8~K~$\leq T \leq$~300~K measured in zero field. The solid curve is the fit by Debye+Einstein models of lattice heat capacity plus crystal field contribution as deduced from inelastic neutron scattering (INS) data. Upper inset: Expanded view of \mbox{low-$T$} $C_{\rm p}(T)$ data over $1.8~{\rm K} \leq T \leq 15$~K\@. Lower inset: Magnetic contribution to heat capacity $C_{\rm mag}(T)$ of Pr$_2$Hf$_2$O$_7$. The solid curve represents the crystal-field contribution to heat capacity according to the CEF level scheme deduced from the INS data.} 
\end{figure}

The $C_{\rm p}(T)$ data of polycrystalline Pr$_2$Hf$_2$O$_7$ are shown in Fig.~\ref{fig:HC} for 1.8~K~$\leq T \leq$~300~K\@. As can be seen from Fig.~\ref{fig:HC} the $C_{\rm p}(T)$ data do not show any anomaly at $T\geq1.8$~K, though an upturn feature is seen at $T \leq 10$~K which is shown in detail in the upper inset of Fig.~\ref{fig:HC}. The magnetic contribution to heat capacity $C_{\rm mag}(T)$ estimated after subtracting the lattice contribution (equivalent to formula mass and unit cell volume corrected $C_{\rm p}(T)$ of La$_2$Hf$_2$O$_7$) is shown in the lower inset of Fig.~\ref{fig:HC}. A broad Schottky type peak near 50~K is seen in $C_{\rm mag}(T)$ data. The $T$ dependence of $C_{\rm mag}$ is described by the crystal field. The crystal field contribution to heat capacity $C_{\rm CEF}(T)$ calculated according to the CEF level scheme deduced from the analysis of INS data (Sec.~\ref{INS}), shown by solid red curve in the lower inset of Fig.~\ref{fig:HC}, agrees well with the $C_{\rm mag}(T)$. 

The room temperature value of  $C_{\rm p} \approx 228$ J/mol\,K is substantially lower than the expected high-$T$ limit Dulong-Petit value, reflecting the large value of Debye temperature $\Theta_{\rm D}$. We estimate $\Theta_{\rm D}$ by fitting the $C_{\rm p}(T)$ data by a combination of the Debye and Einstein models of lattice heat capacity added to the fixed crystal field contribution $C_{\rm CEF}(T)$. The fit of $C_{\rm p}(T)$ data by CEF+Debye+Einstein models \cite{Anand2015a} is shown by the solid red curve in Fig.~\ref{fig:HC} which gives $\Theta_{\rm D} = 790(7)$~K and Einstein temperature $\Theta_{\rm E} = 158(2)$~K with 66\% weight to Debye term and 34\% to Einstein term.  The obtained $\Theta_{\rm D}$ is similar to that of the related pyrohafnates La$_2$Hf$_2$O$_7$ ($\Theta_{\rm D} = 792(5)$~K) and Nd$_2$Hf$_2$O$_7$ ($\Theta_{\rm D} = 785(6)$~K) \cite{Anand2015}. 

\begin{figure*}
\includegraphics[width=\textwidth, keepaspectratio]{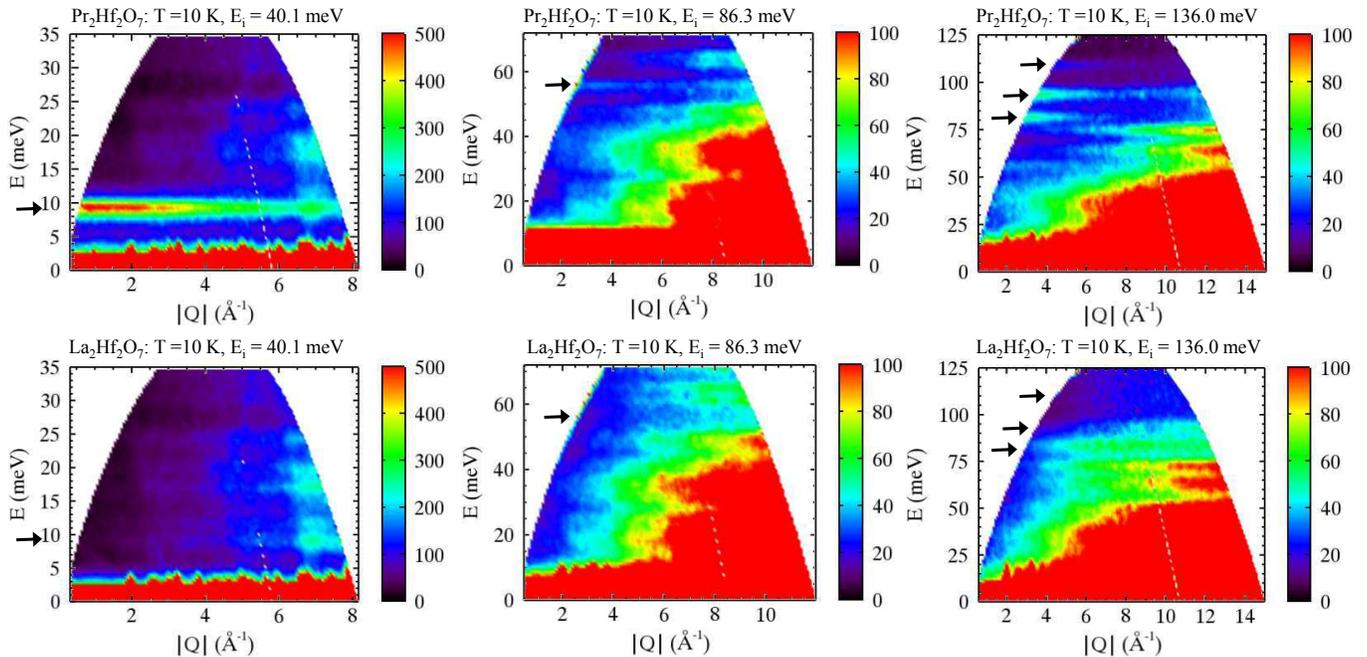}
\caption {(color online) Inelastic neutron scattering response, a color-coded map of the intensity, energy transfer ($E$) versus momentum transfer ($Q$) for Pr$_2$Hf$_2$O$_7$ (upper panel)  and La$_2$Hf$_2$O$_7$ (lower panel) measured at 10~K with the incident energy $E_i=40$~meV, 86~meV and 136~meV. The arrow mark the position of CEF excitations in Pr$_2$Hf$_2$O$_7$. The corresponding energy positions in La$_2$Hf$_2$O$_7$ are also marked. }
\label{fig:INS1}
\end{figure*}

\section{\label{INS} Inelastic Neutron Scattering and Crystal Field Excitations}

The bulk properties measurement discussed above, particularly the ac susceptibility data hint for spin-ice like behavior. Therefore in order to check if the ground state is consistent with the physics of spin-ice which requires a doublet as ground state we determine the crystal field states using inelastic neutron scattering. Figure~\ref{fig:INS1} shows the color-coded contour maps of the INS response from powder Pr$_2$Hf$_2$O$_7$ and La$_2$Hf$_2$O$_7$ showing energy transfer $E$ versus wave vector $Q$ for neutrons of incident energy $E_i=40.1$~meV, 86.3~meV and 136.0~meV at 10~K\@. A comparison of the contour maps of Pr$_2$Hf$_2$O$_7$ and La$_2$Hf$_2$O$_7$ clearly show five dispersionless excitations in Pr$_2$Hf$_2$O$_7$ at around 9.1~meV, 56.5~meV, 82.2~meV, 94.5~meV and 109.5~meV at low-$Q$. Since La$_2$Hf$_2$O$_7$ does not show strong phonon scattering at low-$Q$, these low-$Q$ excitations of Pr$_2$Hf$_2$O$_7$ seems to have magnetic origin due to the crystal field excitations from  Pr$^{3+}$. 

With a $J$ value of 4, the ground state (GS) of Pr$^{3+}$ possesses nine-fold degeneracy, when this $^3H_4$ GS multiplet is subject to a CEF interaction with $D_{3d}$ symmetry (point symmetry $\bar{3}m$) due to the pyrochlore structure, it splits into three doublets and three singlets, designated by the symmetry decompositions 3$\Gamma_3^+ + 2 \Gamma_1^+ + \Gamma_2^+$ in the irreducible representations of $D_{3d}$. Accordingly there ought to be five CEF excitations from the GS multiplet which we clearly see in the INS spectra at 10~K\@. For a $D_{3d}$ symmetry with $z$ axis along the local cubic $\langle$111$\rangle$ direction, the CEF Hamiltonian is given by 
\begin{equation}
\begin{split}
H_{\rm{CEF}}= & B_0^2C_0^2+B_0^4C_0^4+B_3^4(C_{-3}^4+C_3^4)\\
             & +B_0^6C_0^6+B_3^6(C_{-3}^6+C_3^6)+B_6^6(C_{-6}^6+C_6^6),
\end{split}
\end{equation}
where $B_q^k$ are the crystal field parameters and $C_q^k$ the tensor operators \cite{cfbook}. 

\begin{figure}
\includegraphics[width=\columnwidth, keepaspectratio]{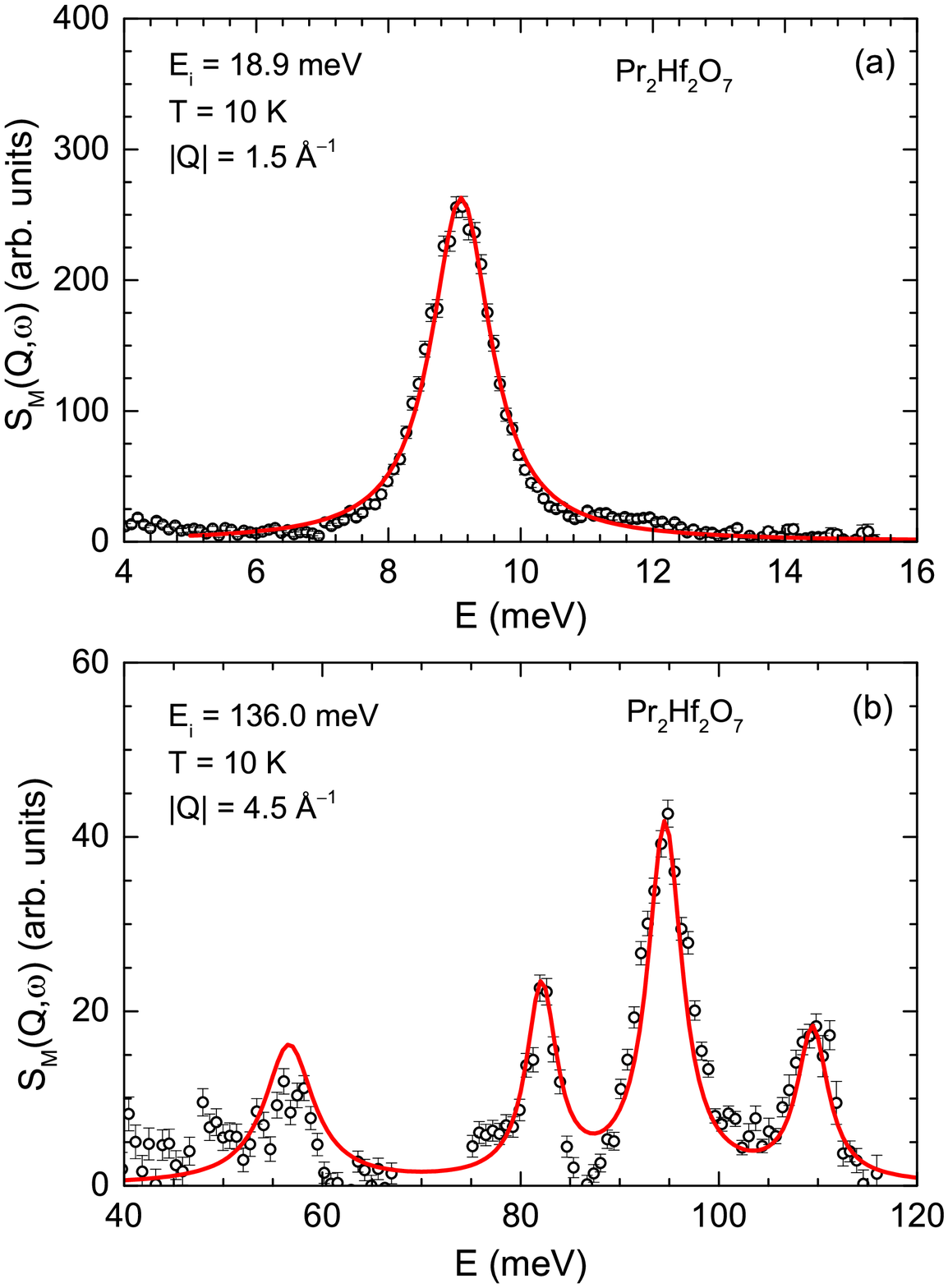}
\caption {(Color online) $Q$-integrated inelastic magnetic scattering intensity $S_{\rm M}(Q,\omega)$ versus energy transfer $E$ of Pr$_2$Hf$_2$O$_7$ integrated over (a) momentum $Q$ range [2.5--3]~{\AA $^{-1}$} for $E_i=18.9$~meV, and (b) $Q$ range [4--5]~{\AA $^{-1}$} for $E_i=136.0$~meV, showing all five transitions within the ground state multiplet $^3H_4$ of Pr$^{3+}$ at 10~K\@. The solid lines are the fits of the data according to the crystal field model discussed in the text.}
\label{fig:INS2}
\end{figure}

The $Q$-integrated one-dimensional energy cuts of the INS response from the low-$Q$ region (where the phonon contributions are relatively weaker) were made to obtain the scattering intensity $S(Q,\omega)$. The magnetic scattering $S_{\rm M}(Q,\omega)$ from Pr$_2$Hf$_2$O$_7$ was obtained by subtracting the phonon contribution using the INS data of nonmagnetic reference La$_2$Hf$_2$O$_7$. The plots of $S_{\rm M}(Q,\omega)$ versus $E$ are shown in Fig.~\ref{fig:INS2} where $S_{\rm M}(Q,\omega)= S(Q,\omega)_{\rm Pr_2Hf_2O_7} - \alpha\, S(Q,\omega)_{\rm La_2Hf_2O_7} $ with $\alpha=0.8$ being the ratio of the neutron scattering cross sections of Pr$_2$Hf$_2$O$_7$ and La$_2$Hf$_2$O$_7$. The five CEF excitations near 9.1~meV, 56.5~meV, 82.2~meV, 94.5~meV and 109.5~meV are clearly visible in the $S_{\rm M}(Q,\omega)$ plots in Fig.~\ref{fig:INS2}. 

\begin{figure}
\includegraphics[width=\columnwidth, keepaspectratio]{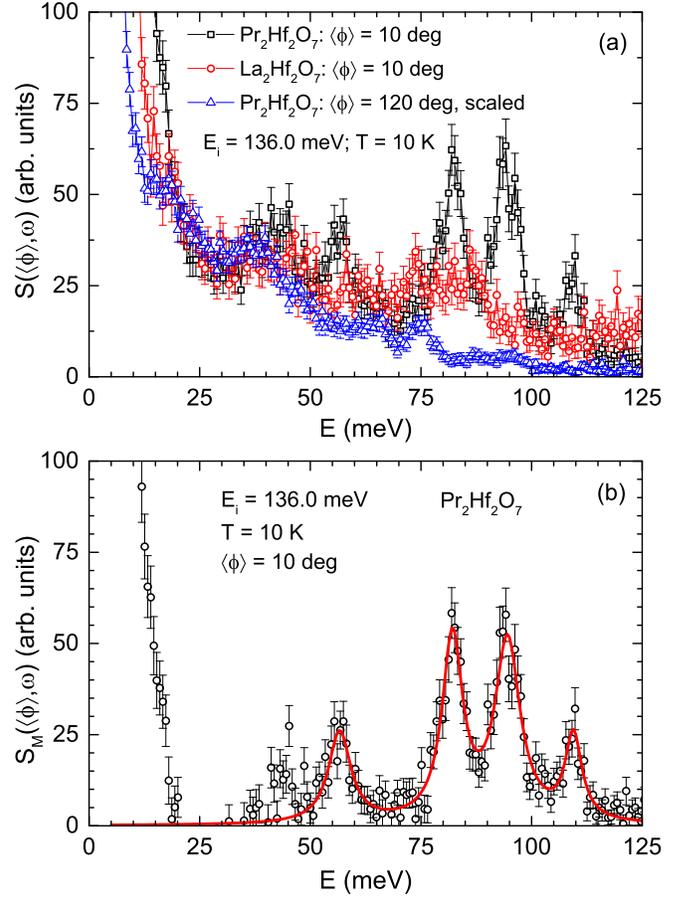}
\caption {(Color online) (a) $\phi $-integrated inelastic scattering intensity $S(\langle \phi \rangle,\omega)$ versus energy transfer $E$ of Pr$_2$Hf$_2$O$_7$ and La$_2$Hf$_2$O$_7$  for $E_i=136.0$~meV integrated over scattering angle  $9^\circ \leq \phi \leq 11^\circ$ together with the scaled phonon backgound $S(\langle \phi \rangle,\omega)$ from $\langle \phi \rangle$ range $119^\circ \leq \phi \leq 121^\circ$ of Pr$_2$Hf$_2$O$_7$. (b) Magnetic scattering intensity $S_{\rm M}(\langle \phi \rangle,\omega)$ versus $E$ of Pr$_2$Hf$_2$O$_7$ for $\langle \phi \rangle$ range $9^\circ \leq \phi \leq 11^\circ$, showing four transitions within the ground state multiplet $^3H_4$ of Pr$^{3+}$ at 10~K\@. The solid line is the fit of the data according to the crystal field model discussed in text. The apparent peak near 45~meV is an artifact due to phonon.}
\label{fig:INS3}
\end{figure}

The $S(Q,\omega)$ of La$_2$Hf$_2$O$_7$ does not account well for the phonon background of Pr$_2$Hf$_2$O$_7$, particularly at low $E$ the phonon contributions to the INS response of the two compounds are quite different. This aspect of the INS data is evident in the $\phi$-integrated energy cuts $S(\langle \phi \rangle,\omega)$ shown in Fig.~\ref{fig:INS3}(a). For a better estimate of the phonon background we adopted the approach suggested by Boothroyd {\it et al}.\ \cite{Boothroyd2003}. Accordingly the low-angle phonon background is estimated from the high-angle scattering of Pr$_2$Hf$_2$O$_7$ itself where the magnetic scattering is negligibly small. For this the high-angle Pr$_2$Hf$_2$O$_7$ spectrum was scaled by a factor equal to the ratio of the high- to low-angle La$_2$Hf$_2$O$_7$ spectra. The scaled phonon background estimated this way from the Pr$_2$Hf$_2$O$_7$ data in $119^\circ \leq \phi \leq 121^\circ$ is shown in Fig.~\ref{fig:INS3}(a). A substantial difference between the phonon background from the La$_2$Hf$_2$O$_7$ spectrum and scaled high-angle Pr$_2$Hf$_2$O$_7$ spectrum is obvious from Fig.~\ref{fig:INS3}(a). The $\phi$-integrated magnetic scattering $S_{\rm M}(\langle \phi \rangle,\omega)$ estimated using this scaled background is shown in Fig.~\ref{fig:INS3}(b). The corresponding $Q$ is obtained using the relation $\hbar^2 Q^2/2m_{\rm n} = E_i + E_f - 2(E_iE_f)^{1/2} \cos\phi$, where $E_f = E_i -E$ and $m_{\rm n}$ is the rest mass of neutron.

In order to obtain quantitative information of the CEF states we analyzed the magnetic INS data using the software SPECTRE \cite{SPECTRE} which allowed us to use the intermediate-coupling free ion basis states for diagonalizing the $H_{\rm CEF}$. We employed the complete set of 91 intermediate coupling basis states following the approach used for the related Pr-based system Pr$_2$Sn$_2$O$_7$ \cite{Princep2013}. For least-square fitting of the observed INS excitations (energies and relative intensities) the CEF parameters of Pr$_2$Sn$_2$O$_7$ were used as the starting parameters. The results of our analysis of INS data are summarized in Table~\ref{tab:cf}. The fits of INS data are shown in Figs.~\ref{fig:INS2}(a), \ref{fig:INS2}(b) and \ref{fig:INS3}(b). We used a Lorentzian shape peak function for fitting the magnetic INS specta. Further details of our analysis can be found in our recent paper on Nd$_2$Zr$_2$O$_7$ \cite{Xu2015} where we employed a similar approach for analyzing the INS data by the crystal field model.

\begin{table}
\caption{\label{tab:cf} Observed  and calculated  crystal-field transition energies ($E$) and integrated intensities ($I$) within the ground state multiplet $^3H_4$ of Pr$_2$Hf$_2$O$_7$ at 10~K\@. The intensities are relative to the highest observed peak. The best-fit crystal field parameters $B_q^k$  are:  $B^2_0 = 33.8$~meV, $B^4_0 = 426.0 $~meV, $B^4_3 = 192.7 $~meV, $B^6_0 = 165.3 $~meV, $B^6_3 = -111.8 $~meV, and $B^6_6 = 178.6$~meV. The goodness-of-fit parameter $\chi^2=3.12$. } 
\begin{ruledtabular}
\begin{tabular}{ccccc}
Levels         & $E_{obs}$ (meV)& $E_{cal}$ (meV)&$I_{obs}$&$I_{cal}$\\
\hline
$\Gamma_3^+$    &  0            & 0        &            & \\
$\Gamma_1^+$    & 9.1(1)      & 8.82   &   -          & - \\
$\Gamma_3^+$    & 56.5(2)    & 56.63  & 0.42(10)     & 0.55\\
$\Gamma_1^+$    & 82.2(3)    & 82.11  & 0.39(8) &0.47\\
$\Gamma_3^+$    & 94.5(3)    & 94.56  & 1.0      &1.0\\
$\Gamma_2^+$    & 109.5(4)  & 109.46 & 0.33(6) & 0.38\\
\end{tabular}
\end{ruledtabular}
\end{table}

The CEF parameters $B_q^k$ obtained from the analysis are: $B^2_0 = 33.8$~meV, $B^4_0 = 426.0 $~meV, $B^4_3 = 192.7 $~meV, $B^6_0 = 165.3 $~meV, $B^6_3 = -111.8 $~meV, and $B^6_6 = 178.6$~meV. The CEF analysis of INS data reveal the ground state to be a doublet, first excited state a singlet at 8.82~meV, second excited state a doublet at 56.63~meV, third excited state a singlet at 82.11~meV,  fourth excited state a doublet at 94.56~meV and fifth excited state a singlet at 109.46~meV. The CEF level scheme of Pr$_2$Hf$_2$O$_7$ is found to be very similar to that of Pr$_2$Sn$_2$O$_7$ \cite{Princep2013} and Pr$_2$Zr$_2$O$_7$ \cite{Kimura2013}.

The wavefunction of the ground state non-Kramers doublet is found to be
\begin{subequations}
\begin{equation}
\begin{split}
\Gamma_{3}^+  = & \ 0.823|^3H_{4},\pm 4 \rangle \pm 0.513|^3H_{4}, \pm 1 \rangle - 0.136|^3H_{4},\pm2 \rangle \\ 
                 &       + 0.139|^1G_{4}, \pm 4 \rangle \pm 0.087|^1G_{4}, \pm 1 \rangle   \\ 
                &      \mp 0.058|^3H_{5},\pm 4 \rangle    \mp 0.050|^3H_{5},\pm 2 \rangle \\ 
\end{split}
\end{equation}
and that of the first excited singlet 
\begin{equation}
\begin{split}
\Gamma_1^+   =\  & 0.173|^3H_{4}, 3 \rangle + 0.950|^3H_{4}, 0 \rangle - 0.173|^3H_{4}, -3 \rangle \\  
               &  + 0.162|^1G_{4}, 0 \rangle + 0.038|^3F_2, 0 \rangle. 
\end{split}
\end{equation}
\label{eq:wavefunctions}
\end{subequations}
A significant mixing of $|^3H_{4},\pm 4 \rangle$ with $|^3H_{4},m_J\neq\pm4 \rangle$ terms as well as with $^1G_{4}$ and $^3H_{5}$ is found in the ground state. Based on the CEF ground state wavefunction we estimate $g_\parallel = 4.80$ and $g_\perp =0$ which are consistent with the values deduced from the magnetic measurements. The ground state wave function gives the magnetic moment of $2.4~\, \mu_{\rm B} $/Pr which is close to the effective moment obtained from the Curie-Weiss analysis of magnetic susceptibility. 

The crystal field parameters obtained from the analysis of INS data are able to reproduce the powder dc susceptibility as well as the magnetic contribution to the heat capacity. The magnetic susceptibility calculated from these CEF parameters is shown in Fig.~\ref{fig:MT}, which show an excellent agreement with the experimental data and thus supports the obtained CEF states and parameters. The anisotropic CEF suceptibility yields $\chi_\parallel/\chi_\perp \sim 45 $ at 10~K where $\chi_\parallel$ is the susceptibility parallel to $\langle 111 \rangle$ and $\chi_\perp$ is perpendicular to $\langle 111 \rangle$. This value of $\chi_\parallel/\chi_\perp $ is very close to that of Pr$_2$Sn$_2$O$_7$ for which $\chi_\parallel/\chi_\perp \sim 60 $ at 10~K \cite{Princep2013}. The $\chi_\perp$ is purely Van-Vleck like therefore $\chi_\parallel/\chi_\perp $ increases very rapidly as $T \rightarrow 0$, e.g.\ at 3.5~K  $\chi_\parallel/\chi_\perp \sim 120 $. The $C_{\rm CEF}(T)$ calculated from the CEF parameters is shown in the lower inset of Fig.~\ref{fig:HC} (solid red line), in good agreement with $C_{\rm mag}(T)$ data, thus further supporting the deduced CEF level scheme. Our CEF level scheme (CEF parameters, eigenvalues and wavefunction) are in good agreement with the recent results by Sibille {\it et al}.\ \cite{Sibille2016}, though the values of observed CEF excitations and deduced $B_q^k$ parameters are slightly different.

\section{\label{PrHfO_SC} Single crystal P\lowercase{r}$_2$H\lowercase{f}$_2$O$_7$ }

\begin{figure}
\includegraphics[width=\columnwidth, keepaspectratio]{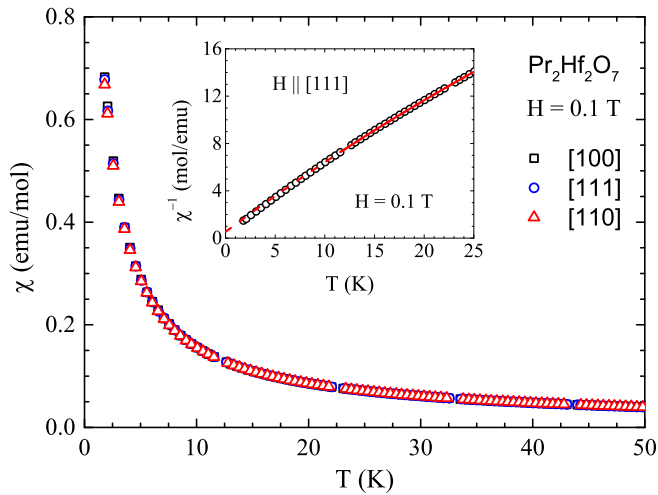}
\caption{(Color online) Zero-field-cooled magnetic susceptibility $\chi$ of single crystal Pr$_2$Hf$_2$O$_7$ as a function of temperature $T$ for $2~{\rm K} \leq T \leq 50$~K for the three crystallographic directions [100], [110] and [111], measured in magnetic field $H= 0.1$~T\@. Inset: Inverse susceptibility $\chi^{-1}(T)$ for $H=0.1$~T and $2~{\rm K} \le T \leq 25$~K\@. The solid red line is the fit of the $\chi^{-1}(T)$ data by the modified Curie-Weiss law in $12~{\rm K} \leq T \leq 25$~K and the dashed line is an extrapolation.}
\label{fig:MT_SC}
\end{figure}

The ZFC $\chi(T)$ of single crystal Pr$_2$Hf$_2$O$_7$ is shown in Fig.~\ref{fig:MT_SC} for the $H$ applied along the three crystallographic directions [100], [110] and [111]. Consistent with the $\chi(T)$ of the polycrystalline sample the $\chi(T)$ of the single crystal does not show any anomaly. The $\chi(T)$ for $H$ along [100], [110] and [111] show very similar $T$ dependence with similar magnitude. In order to estimate the $\theta_{\rm p}$ and $\mu_{\rm eff}$ the $\chi(T)$ data at $12~{\rm K} \leq T \leq 25$~K were analyzed by modified Curie-Weiss behavior, the representative fit for [111] direction is shown in the inset of Fig.~\ref{fig:MT_SC}). The fit parameters are listed in Table~\ref{tab:CW_chi}. The $C$ values yield $\mu_{\rm eff} \approx 2.51\, \mu_{\rm B}$/Pr for the single crystal which, as expected, is very similar to that obtained for the polycrystalline sample. 

\begin{table}
\caption{\label{tab:CW_chi} Parameters obtained from the analysis of magnetic susceptibility of Pr$_2$Hf$_2$O$_7$ by modified Curie-Weiss law. } 
\begin{ruledtabular}
\begin{tabular}{cccc}
$H$ direction          & $\chi_0$            &  $C$                   & $\theta_{\rm p}$\\
                             &  (emu/mol\,Pr)   & (emu\,K/mol\,Pr) &  (K) \\
\hline
$[100]$    & $4.9(1)   \times 10^{-3}$      & 0.79(1)      &   $-0.81(4)$     \\
$[110]$    & $4.8(1)   \times 10^{-3}$      & 0.79(1)      &   $-0.93(5)$      \\
$[111]$    &  $4.9(1)   \times 10^{-3}$     & 0.79(1)      &   $-0.86(5)$     \\
\end{tabular}
\end{ruledtabular}
\end{table}

\begin{figure}
\includegraphics[width=\columnwidth, keepaspectratio]{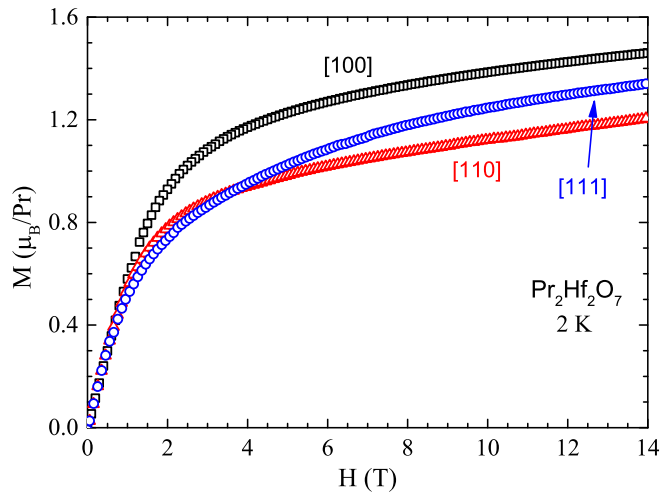}
\caption{(Color online) Isothermal magnetization $M$ of single crystal Pr$_2$Hf$_2$O$_7$ as a function of applied magnetic field $H$ for $0 \leq H \leq 14$~T measured at 2~K for the three crystallographic directions [100], [110] and [111].}
\label{fig:MH_SC}
\end{figure}

The isothermal $M(H)$ data of single crystal Pr$_2$Hf$_2$O$_7$ measured at 2~K are shown in Fig.~\ref{fig:MH_SC}. The $M(H)$ data reflect the Ising anisotropy. For a system with $\langle 111 \rangle$ Ising anisotropy, the saturation magnetizations $M_s $ are given by \cite{Fukazawa2002}: $g' J' \mu_{\rm B}(1/\sqrt 3)$ for [100] (two-in/two-out), $g' J' \mu_{\rm B}(\sqrt{2/3} \times2)/4$ for [110] (one-in/one-out, two free) and $g' J' \mu_{\rm B}(1+1/3\times 3)/4$ for [111] (three-in/one-out). The observed magnetizations at $H=14$~T are $M_{[100]} = 1.46\,\mu_{\rm B}$/Pr, $M_{[110]} = 1.21 \,\mu_{\rm B}$/Pr and $M_{[111]} = 1.34 \,\mu_{\rm B}$/Pr. The observed anisotropic $M$ gives the ratio  $M_{[100]}/M_{[110]}$ = 1.21  and $M_{[100]}/M_{[111]}$ = 1.09  which are in line with Ising anisotropy but somewhat smaller than the theoretically expected ratios of $\sqrt2$ and $2/\sqrt3$, respectively. With $g' J' \mu_{\rm B} \approx 2.51\, \mu_{\rm B} $, the effective moment, one would estimate $M_{[100]} = 1.45\,\mu_{\rm B}$/Pr, $M_{[110]} = 1.02 \,\mu_{\rm B}$/Pr and $M_{[111]} = 1.26 \,\mu_{\rm B}$/Pr. Thus we see that while the observed $M_{[100]} $ and $M_{[111]}$ are close to their expected values, the observed $M_{[110]}$ is significantly higher than the expected value. The deviations from the Ising picture possibly indicates the presence of additional non-Ising terms in the Hamiltonian.  Further we also notice that the $M$ is not saturated even at 14~T field which could be due to the mixing of ground state multiplet and/or non-Ising contributions.

\section{\label{Conclusion} Summary and Conclusions}

We have synthesized polycrystalline and single crystal samples of the pyrohafnate compound Pr$_2$Hf$_2$O$_7$ and investigated  their structural and physical properties by means of synchrotron x-ray diffraction, dc magnetic susceptibility, isothermal magnetization, ac susceptibility, heat capacity and inelastic neutron scattering measurments. The high-resolution synchrotron x-ray diffraction revealed the well-ordered pyrochlore ($Fd\bar{3}m$) structure with Pr/Hf site mixing and/or oxygen deficiency (if present) at a level lower than 0.5\%. Our ac susceptibility measurments down to 90~mK did not reveal any evidence for a magnetic transition to a long range ordered phase. The $\chi_{\rm ac}(T)$ is found to show a frequency dependent broad peak, a characteristic feature of spin-freezing in spin-ice systems, indicating slow spin dynamics in Pr$_2$Hf$_2$O$_7$ at low temperature.  The single-spin relaxation time for Pr$_2$Hf$_2$O$_7$ is found to be very similar to that of the spin-ice Ho$_2$Ti$_2$O$_7$. 

The dc magnetic susceptibility and magnetization data show the Ising anisotropic features commonly observed in 227 pyrochlore spin-ice systems. The single crystal magnetic data clearly reveal this Ising anisotropic behavior, which is most obvious from the field dependence of directional dependent isothermal magnetization data. The inelastic neutron scattering data reveal five well defined magnetic excitations which are well accounted for by a model based on crystal electric field, accordingly we have determined the crystal field states of Pr$^{3+}$. The CEF ground state is a well isolated non-Kramers doublet. The CEF suceptibility yields $\chi_\parallel/\chi_\perp \sim 120 $ at 3.5~K revealing an Ising type anisotropy. The ground state wavefunction shows significant mixing of $|^3H_{4},\pm 4 \rangle$ state with other multiplets. Such a mixing can cause quantum fluctuations. The single crystal $M(H)$ data reflect the presence of non-Ising contribution which are very favorable for quantum spin-ice behavior. 

The key observations from our investigations are i) a weakly ferromagnetic interaction as inferred from the Weiss temperature, ii) crystal field ground state is a doublet well-separated from the first excited state, iii) low-$T$ magnetism described by an effective pseudospin-1/2 model,   iv) anisotropic g-tensor, v) Ising-type anisotropy, and vi) slow spin dynamics. All these indicate that the magnetic ground state of Pr$_2$Hf$_2$O$_7$ is consistent with spin-ice physics. In addition, there is some indication for the presence of a non-Ising contribution that would imply a dynamic spin-ice behavior instead of classical spin-ice behavior. The magnetic properties of Pr$_2$Hf$_2$O$_7$ are very similar to those of the dynamic spin-ice systems Pr$_2$Sn$_2$O$_7$  \cite{Zhou2008,Princep2013} and Pr$_2$Zr$_2$O$_7$ \cite{Kimura2013}. Recent investigations by Sibille {\it et al}.\ [29] also concludes in favor of a quantum spin-ice behavior in Pr$_2$Hf$_2$O$_7$ where they also implied low energy inelastic neutron scattering in addtion to ac susceptibility for the study of spin dynamics in this compound. Further investigations are required to explore the quantum spin-ice ground state of Pr$_2$Hf$_2$O$_7$.

\acknowledgements
We acknowledge Helmholtz Gemeinschaft for funding via the Helmholtz Virtual Institute (Project No. VH-VI-521) and DFG through Research Training Group GRK 1621 and SFB 1143. We also acknowledge the support of the HLD-HZDR, a member of the European Magnetic Field Laboratory (EMFL).

\end{document}